\documentclass[12pt,a4paper]{article}

\usepackage[margin=3.5cm]{geometry}
\usepackage{amsmath,amssymb,cancel}
\usepackage{stackrel}
\usepackage{graphicx,hyperref}
\usepackage{epstopdf}
\usepackage{xcolor}
\usepackage{slashed}
\usepackage{float}
\usepackage{booktabs}
\usepackage[T1]{fontenc}
\usepackage[utf8]{inputenc}
\usepackage{authblk}
\usepackage[bottom]{footmisc}
\usepackage{enumitem}

\definecolor{curve}{rgb}{1,.4,.3}

\newcommand{\curve}[1]{\textcolor{curve}{#1}}

\numberwithin{equation}{section}

\begin{document}


\thispagestyle{empty}

\vspace*{1cm}

\begin{center}
	
	{\Large \bf A New Example of the Effects of a Singular Background on the Zeta Function \par}
	
	\vspace{1cm}
	
	{\large
		Horacio Falomir,
		Joaquín Liniado,
		Pablo Pisani}\\\vspace{15mm}
	\textit{Instituto de F\'isica La Plata (IFLP),\\ CONICET and Universidad Nacional de La Plata (UNLP),\\ CC 67 (1900) La Plata, Argentina
	\footnote{\scriptsize\texttt{falomir@fisica.unlp.edu.ar; joacoliniado@hotmail.com; pisani@fisica.unlp.edu.ar}}}\\\vspace{10mm}
	This article is dedicated to the memory of our dear friend and colleague María Amelia Muschietti.
	
\end{center}

\vspace{10mm}

\begin{abstract}
	\noindent To motivate our discussion, we consider a $1+1$ dimensional scalar field interacting with a static Coulomb-type background, so that the spectrum of quantum fluctuations is given by a second-order differential operator on a single coordinate $r$ with a singular coefficient proportional to $1/r$. We find that the spectral functions of this operator present an interesting behavior: the $\zeta$ function has multiple poles in the complex plane; accordingly, logarithms of the proper time appear in the heat-trace expansion. As a consequence, the $\zeta$ function does not provide a finite regularization of the effective action. This work extends similar results previously derived in the context of conical singularities.
\end{abstract}

\pagebreak

\tableofcontents


\section{Introduction}

Spectral functions---such as the $\zeta$ function and the heat-trace---encode information on the quantum properties of a field theory. Among the various uses of spectral functions in QFT is the computation of functional determinants. For example, in a local field theory the one-loop effective action is given by the functional determinant of a certain differential operator $D$, that can be computed as
\begin{align}\label{fundet}
	\log{\rm Det}\,D=-\zeta'(0)\,,
\end{align}
where $\zeta(s)={\rm Tr}\,D^{-s}$ is the $\zeta$ function of the operator $D$. This definition of the determinant---originally proposed in \cite{Dowker:1975tf,Hawking:1976ja} for QFT; previously in \cite{Ray-Singer} in a geometrical context---assumes that $\zeta(s)$ is analytic at $s=0$. For this property to hold, one demands that the coefficients of the differential operator $D$ are well-behaved functions on a smooth manifold $M$---either without \cite{Seeley1} or with boundaries \cite{Seeley2,Seeley3}. As a matter of fact, under these regularity and smoothness conditions, $\zeta(s)$ admits a meromorphic extension to the whole complex plane $s\in\mathbb{C}$ with, possibly, isolated simple poles at $s_k=\frac{d-k}{n}\notin\mathbb{Z}^-$, where $d$ is the dimension of $M$, $n$ is the order of $D$, and $k$ is a positive integer \cite{Gilkey:1995mj}. Some applications rely on the fact that the residues contain information on physical quantities of the field theory.

Nevertheless, many interesting models in QFT involve quantum fields in the presence of singularities, either at the base manifold $M$ or at the coefficients of $D$. Thus, a natural question arises as to whether this well-known meromorphic structure of the $\zeta$ function persists even in the presence of singular backgrounds.

In 1980, C.\ Callias and C.\ H.\ Taubes \cite{Callias:1979fi} studied the functional determinant of a second-order differential operator with singular coefficients on $\mathbb{R}^4$ arising in the context of a fermionic quantum field in the background of Yang-Mills multi-instantons. In appendix A of their article they conjectured\footnote{Actually, their statement refers to the small proper-time asymptotics of the heat-kernel, which we rephrase in terms of the $\zeta$ function.} that if a differential operator has singular coefficients then multiple poles might appear at locations which are not necessarily half-integers, as in the smooth case. However, simple examples which exhibit such behavior were not known for some time.

In 1996, E.\ Mooers provided the first model which, due to the presence of singularities, has a $\zeta$ function with poles at unexpected locations \cite{Mooers1} (see also \cite{Mooers2}). Her analysis is based on the selfadjoint extensions of the Laplacian of conical manifolds, and shows that the $\zeta$ function has simple poles whose locations depend on the deficiency angle at the tip of the cone.

Later, in collaborations with M.~A.~Muschietti, R.~Seeley, and A.~Wipf, two of the authors of the present article obtained similar results in settings where the role of the conical singularity is played by $1/r^2$-type potentials \cite{Falomir:2001iw,Falomir:2003vw,Falomir:2004zf,Falomir:2005yw,Falomir:2005xh} (for a review on these results, see \cite{Falomir:2012rk}). These articles provide a complete description of the analytic structure of the $\zeta$ function. The effect of the singularity in Laplace-type operators of the form $A=-\triangle+\alpha/r^2$ can be understood as follows. The operator is formally scale invariant but, at the same time, requires an appropriate definition of boundary conditions at $r=0$. A standard analysis of the radial part of $A$ shows that there are infinitely many admissible boundary conditions, which are characterized by a single real parameter $\beta$ whose length-dimension is non-integer but depends on $\alpha$. Therefore, the length-dimensions of the Seeley-DeWitt coefficients (viz.\ the location of the poles of the $\zeta$ function) are expected to depend on $\alpha$, too. In other words, the breaking of scale invariance introduced by the boundary condition is responsible for the exotic pole structure of the $\zeta$ function: its meromorphic extension has simple poles at locations determined by the ``external'' parameter $\alpha$. However, Dirichlet boundary conditions (corresponding to $\beta=\infty$), as well as the case $\beta=0$, preserve scale invariance; as a consequence, for these two selfadjoint extensions the poles have the usual location at negative half-integers.

K.\ Kirsten, P.\ Loya and J.\ Park further elaborated on the problem of spectral functions under $1/r^2$-type potentials \cite{Kirsten:2005bh}, focusing on the limit case that determines the threshold beyond which the operator is unbounded from below. They found a remarkable property (which had been overlooked in \cite{Falomir:2004zf}\footnote{See also the detailed comparison with the article by Mooers in \cite{V}.}) that prevents a straightforward application of definition \eqref{fundet}: the $\zeta$ function might have a branch-cut at $s=0$,
\begin{align}\label{branch}
\zeta(s)\simeq s\log{s}+{\rm (holomorphic\ terms)}\,.
\end{align}

Soon after, the same authors established a complete description of the analytic properties of the $\zeta$ function for conical singularities \cite{Kirsten:2005yw,Kirsten:2007ur,Kirsten:2008wu}. In these articles they construct $\zeta$ functions that exhibit new ``pathological'' properties: a simple pole at $s=0$ and poles of increasing multiplicity---as well as logarithmic cuts---at unusual locations. Thus, the $\zeta$ function for conic manifolds does not admit, in general, a meromorphic extension at all. 

Generalizations and applications of these results to more formal settings are vast and include warped cones, non-selfadjoint operators, $\Psi$DO's, functional determinants, analytic torsion, etc.\ (see e.g.\ \cite{GL,GKM,Gil:2009ay,LMP,V1,LV} and references therein). However, up to our knowledge, analogous studies on other types of singularities have not been carried out yet. To this aim, we analyze in the present article the milder Coulomb-type singularity and find two results worth remarking: (i) the $\zeta$ function might have some of the unusual properties found for conical singularities; (ii) this unusual behavior is present even for Dirichlet boundary conditions. As already mentioned, in the case of conical singularities the existence of a family of boundary conditions is essential for the unusual behavior of the $\zeta$ function---for this reason, in many classical studies of $1/r^2$-type singularities the unusual behavior does not show up (see e.g.\ \cite{Cheeger1,Cheeger2,Callias1,Callias2,B-S1,B-S2,B-S3,Lesch}, to mention just a few).

In this article we consider the whole family of selfadjoint extensions for the Coulomb-type operator and find that the $\zeta$ function has infinitely many poles of increasing multiplicity. However, when Dirichlet boundary condition is imposed one does not recover the usual meromorphic extension but a single double pole remains\footnote{This property had already been noted in a preliminary approach in \cite{Pisani:2007cu}}.

Specifically, we study in some detail and from different perspectives the one-dimensional second-order differential operator $A=-\partial^2_r+\alpha/r$. Interestingly, the existence of multiple poles at negative half-integers allows one to construct---through the addition of extra dimensions---an operator whose $\zeta$ function has a simple pole at $s=0$. In this way, we provide a model for which
\begin{align}
	\zeta(s)\simeq \frac{1}{s}+{\rm (holomorphic\ terms)}\,.
\end{align}
As in the model presented in \cite{Kirsten:2005bh}, whose $\zeta$ function satisfies \eqref{branch}, the $\zeta$ function for the Coulomb-type background is not analytic at $s=0$ either, so definition \eqref{fundet} does not provide a finite regularization of the effective action.

This article is organized as follows. In Section \ref{qft} we embed the problem of interest in the context of a quantum field interacting with an external singular background. At the same time, this section introduces the standard procedure for applying spectral functions in the absence of singularities. In Section \ref{theoperator} we analyze the spectrum of the singular operator $A$, whose $\zeta$ function is studied in Section \ref{seczeta}. In Section \ref{theheattrace} we recast our findings for the $\zeta$ function in terms of logarithms in the heat-trace expansion. Section \ref{effact} refers to the consequences of the multiple poles on the computation of the effective action; in particular, we show that the $\zeta$-function regularization of the model introduced in Section \ref{qft} is not finite. Section \ref{bc} contains an analysis of the most general boundary conditions that can be imposed at the singular point $r=0$; since we adopt an intuitive but not rigorous approach, Appendix \ref{sae} relates the results of this section with the formal construction of selfadjoint extensions. Finally, section \ref{higher-poles} is devoted to the computation of the multiple poles for the case of general boundary conditions. To conclude, in Section \ref{conclu} we summarize our results and draw some conclusions.


\section{Quantization on a Classical Background}\label{qft}

This preliminary section is aimed at providing a self-contained review of some aspects of the use of spectral functions in the computation of the effective action due to an interaction with a classical background.

Let us consider a typical quantum field $\varphi(x)$ whose dynamics is given by a Euclidean classical action $S[\varphi]$. For any arbitrary external source $J(x)$, the mean value $\phi(x)$ of the quantum field is determined by the functional integral
\begin{align}
	\phi(x)=\int\mathcal D\varphi
	\ \varphi(x)
	\ e^{-S[\varphi]+\int dx\,J(x)\varphi(x)}\,.
\end{align}
We assume that the source $J(x)$ and the mean field $\phi(x)$ uniquely determine each other. The quantum effective action $\Gamma[\phi]$ can be defined through the average
\begin{align}
	e^{-\Gamma[\phi]+\int dx\,J(x)\phi(x)}=\int\mathcal D\varphi
	\ e^{-S[\varphi]+\int dx\,J(x)\varphi(x)}
\end{align}
for any arbitrary source $J(x)$ or, equivalently, mean field $\phi(x)$. Expanding the integration field $\varphi(x)$ around $\phi(x)$ one readily obtains, at one-loop order,
\begin{align}\label{funder}
	\Gamma[\phi]=S[\phi]
	+\tfrac12\,\log{{\rm Det}\,\frac{\delta^2S[\phi]}{\delta\phi(x)\delta\phi(x')}}\,.
\end{align}
For a local action $S[\phi]$, the second functional derivative produces a differential operator, whose determinant is generally divergent: this is a manifestation of the one-loop divergences of quantum corrections. Apart from the diagrammatic techniques, the heat-kernel and the $\zeta$ function provide two of the most commonly used procedures for regularizing these determinants.

Let us take as an example a real scalar field $\varphi(x)$, with $x^\mu=(t,\vec r)\in\mathbb{R}^{1+3}$, in interaction with a classical static background $V(\vec{r})$. The corresponding Euclidean action is
\begin{align}
	S[\varphi]=\tfrac12\,\partial_\mu\varphi\,\partial_\mu\varphi+\tfrac 12 m^2\,\varphi^2+\tfrac12V(\vec{r})\,\varphi^2\,.
\end{align}
Quantum fluctuations are given by the functional derivative in \eqref{funder}, which corresponds to the second-order differential operator
\begin{align}\label{d}
	D=-\partial^2_t+A+m^2\,,
\end{align}
where
\begin{align}\label{a}
	A=-\triangle+V(\vec r)
\end{align}
is a Laplace-type operator that acts on $\mathbf{L}_2(\mathbb{R}^3)$. An appropriate regularization of the functional determinant can be implemented by means of Schwinger's proper-time procedure, in which the determinant ${\rm Det}\,D$ is written in terms of the heat-trace\footnote{This definition is inspired by the identity $\log{(\lambda/\mu^2)}=-\int_0^\infty d\tau\,\tau^{-1}(e^{-\tau\lambda}-e^{-\tau\mu^2})$.},
\begin{align}\label{effactdiv}
\log{\rm Det}\,D
=-\int_{\Lambda^{-2}}^\infty \frac{d\tau}{\tau}
\ {\rm Tr}\,e^{-\tau\,D}
=-T\int_{\Lambda^{-2}}^\infty \frac{d\tau}{\tau}
\ \frac{e^{-\tau m^2}}{\sqrt{4\pi \tau}}
\ {\rm Tr}\,e^{-\tau\,A}\,.
\end{align}
In this representation UV-divergences are handled through the UV cutoff $\Lambda$ which, eventually, tends to infinity. In the last equality $T$ represents the infinite timelike volume. 

Spectral theory shows that for smooth backgrounds $V(\vec r)$ on $\mathbb{R}^d$, the heat-trace admits the small-$\tau$ asymptotic expansion \cite{Gilkey:1995mj}
\begin{align}\label{htexp}
	{\rm Tr}\,e^{-\tau\,A}\sim
	\frac{1}{(4\pi\tau)^{\frac d2}}\ \sum_{n=0}^\infty a_n(A)\,\tau^{n}\,,
\end{align}
where the Seeley-DeWitt coefficients $a_n(A)$ involve the integral on $\mathbb{R}^d$ of the background $V(\vec r)$ and its derivatives. In a more general setting---such as on curved spacetimes---the Seeley-DeWitt coefficients also include information on the geometric properties of the base manifold. Furthermore, in the presence of boundaries the asymptotic expansion \eqref{htexp} includes half-integer values of $n$ as well.

In this way, representation \eqref{effactdiv} translates the small-$\tau$ behavior of the heat-trace into UV divergences. In fact, replacing \eqref{htexp} (for $d=3$) into \eqref{effactdiv} one isolates the divergent contributions to the one-loop effective action,
\begin{align}\label{divea}
&\Gamma[\phi]=S[\phi]
-\frac{T\,a_0(A)}{64\pi^2}
\left(\Lambda^4-2m^2\Lambda^2+2m^4\log{\left(\tfrac{\Lambda}{m}\right)}\right)
+\mbox{}\nonumber\\[2mm]
&\mbox{}-\frac{T\,a_1(A)}{32\pi^2}
\left(\Lambda^2-2m^2\log{\left(\tfrac{\Lambda}{m}\right)}\right)
-\frac{T\,a_2(A)}{16\pi^2}\,\log{\left(\tfrac{\Lambda}{m}\right)}
+\Gamma_{\rm finite}[\phi]\,.
\end{align}
The term $\Gamma_{\rm finite}[\phi]$ is finite as $\Lambda\to\infty$. For a smooth background $V(\vec r)$, the first Seeley-DeWitt coefficients are well-known \cite{Vassilevich:2003xt}. In particular, for $d=3$,
\begin{align}
	a_0(A)&=\int_{\mathbb{R}^3}d^3r\ \ 1\,,\label{a0}\\[2mm]
	a_1(A)&=-\int_{\mathbb{R}^3}d^3r\ \ V(\vec r)\,,\label{a1}\\[2mm]
	a_2(A)&=\int_{\mathbb{R}^3}d^3r\,
	\left(\tfrac12\,V^2(\vec r)-\tfrac16\,\triangle V(\vec r)\right)\,.\label{a2}
\end{align}
In this way, Seeley-DeWitt coefficients indicate how UV divergences are removed through a renormalization of bare parameters in the action: as shown by \eqref{a0}, the $O(\Lambda^4)$-divergence in \eqref{divea} can be removed through a renormalization of the cosmological constant; on the other hand, the remaining $O(\Lambda^2)$- and $\log{\Lambda}$-divergences in \eqref{divea} depend on the background so are related to the bare parameters involved in the dynamics of the field $V(\vec r)$. The rest of the effective action is then given by $\Gamma_{\rm finite}[\phi]$.

Alternatively, the functional determinant of a differential operator $D$ can also be computed using the $\zeta$ function
\begin{align}\label{zetad}
	\zeta(s)={\rm Tr}\,\left(\frac{D}{\mu^2}\right)^{-s}=\sum_{n=1}^\infty \left(\frac{\lambda_n}{\mu^2}\right)^{-s}
\end{align}
where $\lambda_n$ denote the eigenvalues of $D$. Note we have introduced an arbitrary mass scale $\mu$ for dimensional consistency. This definition must be used for ${\rm Re}(s)$ large enough so that the series converges to an analytic function; otherwise, $\zeta(s)$ is defined as its analytic extension to the rest of the complex plane.

The functional determinant in the $\zeta$ function approach is then defined as
\begin{align}\label{detzet}
	\log{\rm Det}\left(\frac{D}{\mu^2}\right)
	=\sum_{n=1}^\infty \log{\left(\frac{\lambda_n}{\mu^2}\right)}
	=-\zeta'(0)\,.
\end{align}
The series in this expression is not convergent but should be understood as a formal motivation for the last equality. One can prove that the analytic extension of $\zeta(s)$ admits a finite derivative at $s=0$. Indeed, the $\zeta$ function and the heat-trace are related by a Mellin transform\footnote{This relation is a direct consequence of the identity $\Gamma(s)\,\lambda^{-s}=\int_0^\infty d\tau\,\tau^{s-1}\,e^{-\tau\lambda}$.},
\begin{align}\label{mellin}
\zeta(s)=\frac{\mu^{2s}}{\Gamma(s)}
\int_0^\infty d\tau\ \tau^{s-1}\,{\rm Tr}\,e^{-\tau\,A}\,.
\end{align}
This expression\footnote{Note that \eqref{mellin} gives an alternative regularization of the divergent integral in \eqref{effactdiv}: in that case one introduced an UV cutoff whereas the analytic extension suggested by \eqref{mellin} gives instead $-\zeta'(0)$ plus a divergent term proportional to $\zeta(0)$.}, together with the asymptotic expansion \eqref{htexp}, shows that $\zeta(s)$ is analytic at $s=0$: the integral close to $\tau\simeq 0$ has a simple pole at $s=0$ (with residue proportional to $a_{d/2}(A)$) which is canceled by $\Gamma(s)$ in the denominator.

All in all, both the heat-trace and the $\zeta$ function can be used to compute the effective action: the differences between these approaches amount to distinct choices of the renormalization scheme. In particular, the heat-trace produces UV-divergent terms whereas, as just shown, the $\zeta$ function leads---in the smooth background case---to a finite renormalization.

The model we analyze in the present article provides a simple example in which the $\zeta$ function regularization is not finite. Actually, these quite general procedures cannot be straightforwardly applied in the presence of singular backgrounds. In fact, as can be already seen from \eqref{a1} and \eqref{a2}, the well-known expressions for the Seeley-DeWitt coefficients are ill-defined if the backgrounds $V(\vec r)$ (or their powers or derivatives) contain non-integrable singularities.

In this article we analyze a singular Coulomb-type background and find interesting properties of the spectral functions which depart from the standard behavior described in this section: the asymptotic expansion of the heat-trace is not of the form \eqref{htexp} but also contains logarithms of the proper time. Correspondingly, the $\zeta$ function \eqref{zetad} has a simple pole at $s=0$. For simplicity, we just present the analysis corresponding to a field in 1+1 dimensions.

\section{The Singular Operator}\label{theoperator}

In relation to the context set up in the previous section, we consider a scalar field $\varphi(t,r)$ on a compact spacelike segment $r\in[0,L]$ in interaction with a Coulomb-type background
\begin{align}
	V(r)=\frac{\alpha}{r}\,,
\end{align}
where $\alpha$ is a constant with dimension of mass. To compute the corrections introduced by quantum fluctuations we must analyze the spectral properties of the second-order differential operator,
\begin{align}\label{operator}
  A=-\partial_r^2+\frac{\alpha}{r}\,,
\end{align}
acting on square-integrable functions $\varphi(r)\in\mathbf{L}_2(0,L)$. Our purpose is to determine the effect of the singularity on the spectral functions of $A$.

In general, the microscopic interaction of the wave function $\varphi(r)$ with the singularity at $r=0$ is characterized by the boundary condition. In this first part of the article we consider Dirichlet boundary conditions at both endpoints,
\begin{align}\label{diri}
	\varphi(0)=\varphi(L)=0\,,
\end{align}
because this suffices to exhibit some unusual spectral properties of the operator $A$. Afterwards, we will analyze which other boundary conditions can be imposed at the singularity as well as their consequences on the spectral functions.

The eigenvalue equation
\begin{align}\label{equation}
  \left(A-z^2\right)\varphi_z(r)=0\,,
\end{align}
where we assume $z$ belongs to the right half-plane ${\rm Re}(z)\geq 0$, has the following set of (unnormalized) solutions:
\begin{align}
  \psi_z(r)&= r\,e^{-iz r}
  \ M\left(1+\frac{\alpha}{2iz},2;2iz\, r\right)\,,\label{m}\\[2mm]
  \chi_z( r)&= r\,e^{-iz r}
  \ U\left(1+\frac{\alpha}{2iz},2;2iz\, r\right)\,;\label{u}
\end{align}
$M$ and $U$ are confluent hypergeometric functions \cite{A-S}, which are linearly independent for $1+\alpha/2iz \notin \mathbb{Z}^-$. Close to the singularity the independent solutions \eqref{m} and \eqref{u} behave as $\psi_z( r)\sim O( r)$ and $\chi_z( r)\sim O(1)$, so only the former satisfies the boundary condition at the singularity.

Let us first determine the non-positive eigenvalues of $A$. Numerical analysis shows that only one eigenstate with negative eigenvalue exists as long as $\alpha L$ is less than a critical value $-3.67049266\ldots$

As regards zero modes, only for a discrete set of negative values of $\alpha L$, given by $-\frac14\,j_1\mbox{}^2$ (where $j_1$ is any zero of the Bessel function $J_1$) does \eqref{equation} admit a solution with $z=0$. Not surprisingly, the highest value of $\alpha$ for which $A$ has a zero mode is given by the first zero of $J_1$ as $\alpha L=-3.67049266\ldots$

Since we are interested in the growth of the eigenvalues and there is only a finite number of those which are non-positive, in the sequel we restrict for simplicity to the case $\alpha>0$.

Under Dirichlet boundary conditions, both at $ r=0$ and at $ r=L$, the discrete positive spectrum $z^2_n$ of $A$ is given by the solutions to the following transcendental equation:
\begin{align}\label{spectrum}
    e^{-i\,L z_n}M\left(1+\frac{\alpha}{2iz_n},2;2iL\,z_n\right)=0\,.
\end{align}
The exponential in this expression is retained for later convenience. As we will see, the appearance of $z_n$ in both arguments of the confluent hypergeometric function will be crucial for the unusual asymptotic behavior of the eigenvalues.

\begin{figure}[t]
\centering
\begin{minipage}{.46\textwidth}
\includegraphics[height=40mm]{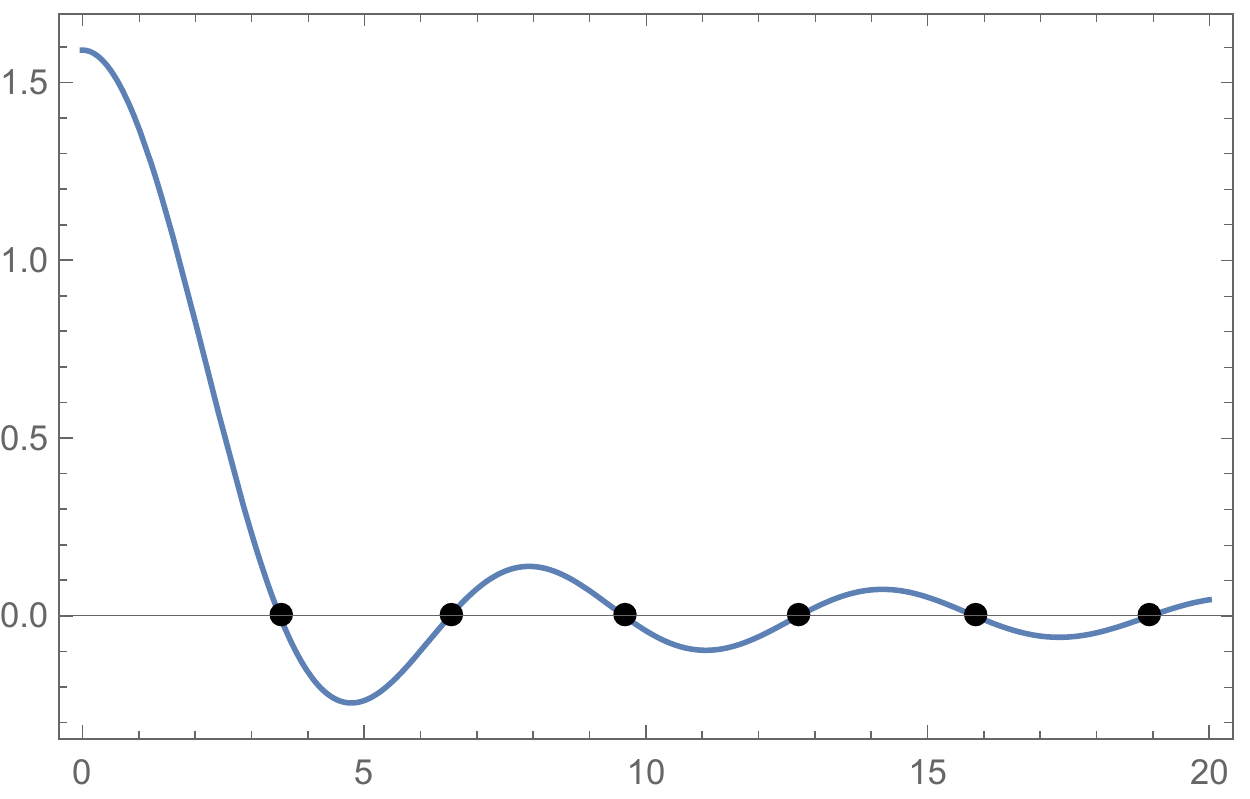}
\caption{\small{The intersections $z_n$ with the positive real axis give the spectrum $z_n^2$ of $A$ for $\alpha=1$. $L$ is taken as the unit length.}}
\label{eigenvalues}
\end{minipage}
\hspace{0.02\textwidth}
\begin{minipage}{.46\textwidth}
\centering
\includegraphics[height=40mm]{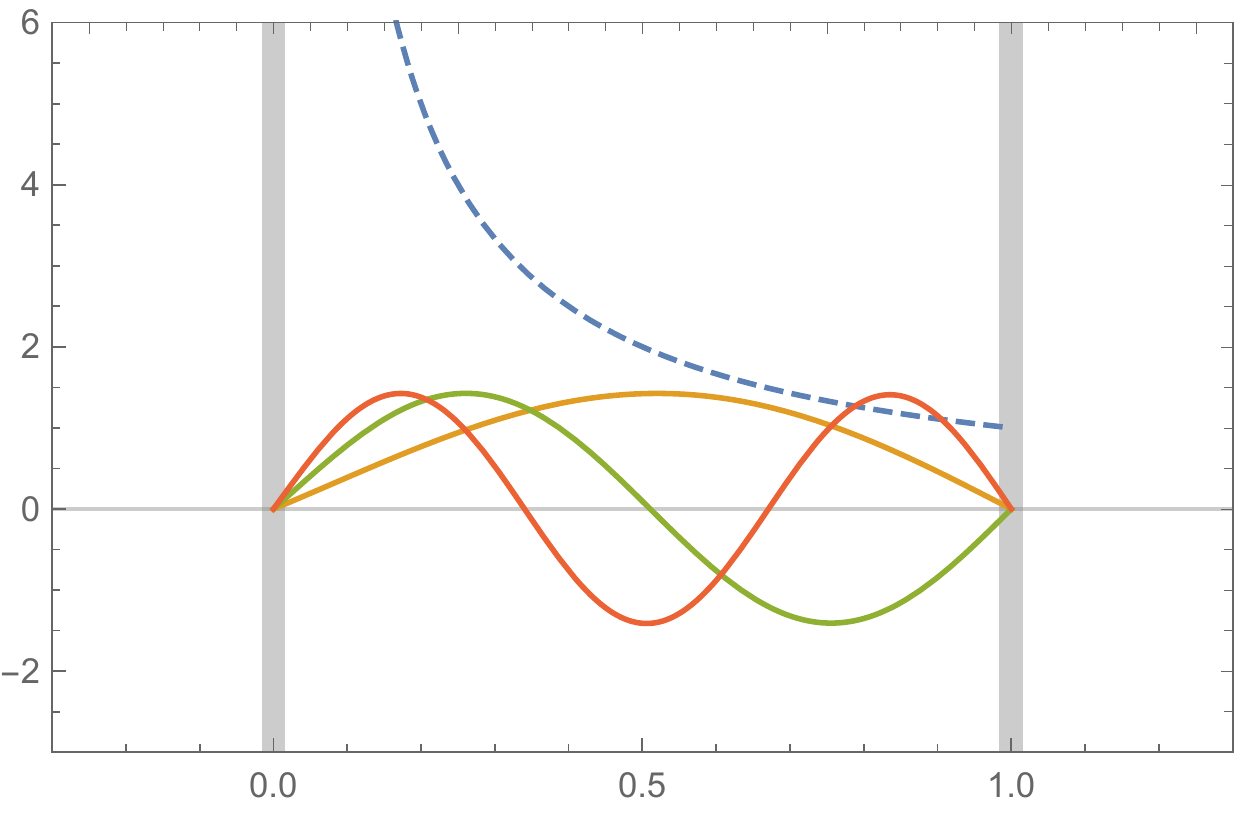}
\caption{\small{Normalized eigenfunctions for $\alpha=1$. The dashed line shows the potential $1/ r$. $L$ is taken as the unit length.}}
\label{eigenfunctions}
\end{minipage}
\end{figure}

Figure \ref{eigenvalues} displays the l.h.s.\ of \eqref{spectrum} as a function of $z_n$, for $\alpha L=1$; the zeroes determine the spectrum of $A$. Figure \ref{eigenfunctions} shows the first three lowest eigenfunctions between the rigid boundaries at $ r=0$ and $ r=L$: the wave functions are (slightly) driven away from the singularity.

\section{The Zeta Function}\label{seczeta}

In this section we analyze the $\zeta$ function of the singular operator $A$ defined on the compact domain of length $L$,
\begin{align}\label{zeta}
	\zeta(s)=\mu^{2s}\ {\rm Tr}\,A^{-s}
	=\mu^{2s}\ \sum_{n=1}^\infty z_n^{-2s}\,.
\end{align}
In particular, we will focus our study on its analytic structure, which is determined by the asymptotic behavior of the spectrum. For large values of $z_n$, \eqref{spectrum} is asymptotically equivalent \cite{A-S} to
\begin{align}
{\rm Re}
\left[\frac{e^{i\left(Lz_n-\frac{\alpha}{2z_n}\log{(2Lz_n)}\right)}}{\Gamma\left(1+i\frac{\alpha}{2z_n}\right)}
\,\sum_{k=0}^\infty
\,\frac{\Gamma\left(k+1+i\frac{\alpha}{2z_n}\right)\Gamma\left(k+i\frac{\alpha}{2z_n}\right)}{k!\,(2iLz_n)^{k}}
\right]\sim 0\,,
\end{align}
which can be recursively solved by expanding $z_n$ for large values of $n$. The result for the leading contributions reads
\begin{align}\label{asymptotic-eigenvalues}
z_n&\sim \frac{\pi\,n}{L}+\frac{\alpha}{2\pi n}\left(\log{(2\pi n)}+\gamma\right)-\mbox{}\nonumber\\[2mm]
&\mbox{}-\frac{\alpha}{(2\pi n)^3}
\ \Big\{2\alpha L\log^2{(2\pi n)}+2(2\gamma-1)\,\alpha L\log{(2\pi n)}+\mbox{}\nonumber\\[2mm]
&\mbox{}+\frac{\zeta_R(3)}{3}\,(\alpha L)^3+(1-2\gamma+2\gamma^2)\,\alpha L-1\Big\}+O(n^{-5})\,,
\end{align}
where $\gamma$ is Euler's constant. The term indicated as $O(n^{-5})$ contains a cubic polynomial in $\log{(2\pi n)}$. This approximation to the large eigenvalues---derived from an asymptotic expansion---is illustrated in Table \ref{asymptotic}, which shows the first 13 values of $z_n$ for $\alpha=L=1$ together with the successive approximations obtained by considering higher-order terms in the asymptotic expansion \eqref{asymptotic-eigenvalues}, including up to the $O(n^{-5})$ correction.
\begin{table}[t]
	\centering
\begin{minipage}{.75\textwidth}
	\scriptsize
	\centering
	{\setlength{\doublerulesep}{1pt}
		\begin{tabular}{@{}cccccc@{}}
			\toprule\toprule
			$O(n)$    & $O(n^{-1})$   & $O(n^{-3})$   &  $O(n^{-5})$   &    &{\rm Eigenvalues}\\[0.5ex]
			\midrule
			\addlinespace[3pt]
			3.141592654 & 3.525966600 & 3.496795708 & 3.502157775 & $\ldots$ & 3.500788704 \\
			6.283185307 & 6.530531180 & 6.523724838 & 6.524144307 & $\ldots$ & 6.524098380 \\
			9.424777961 & 9.611185802 & 9.608488391 & 9.608573947 & $\ldots$ &  9.608568384\\
			12.56637061 & 12.71762300 & 12.71625615 & 12.71628294 & $\ldots$ &  12.71628173\\
			15.70796327 & 15.83606806 & 15.83526998 & 15.83528069 & $\ldots$ &  15.83528033\\
			18.84955592 & 18.96114614 & 18.96063508 & 18.96064010 & $\ldots$ &  18.96063996\\
			21.99114858 & 22.09030217 & 22.08995291 & 22.08995554 & $\ldots$ &  22.08995548\\
			25.13274123 & 25.22215715 & 25.22190665 & 25.22190815 & $\ldots$ &  25.22190812\\
			28.27433388 & 28.35589756 & 28.35571107 & 28.35571198 & $\ldots$ &  28.35571196\\
			31.41592654 & 31.49101071 & 31.49086768 & 31.49086826 & $\ldots$ &  31.49086825\\
			34.55751919 & 34.62715654 & 34.62704415 & 34.62704454 & $\ldots$ &  34.62704453\\
			37.69911184 & 37.76410010 & 37.76401000 & 37.76401027 & $\ldots$ &  37.76401026\\
			40.84070450 & 40.90167360 & 40.90160012 & 40.90160031 & $\ldots$ &  40.90160031\\
			\bottomrule\bottomrule
			\addlinespace[2mm]
	\end{tabular}}
	\caption{\small Successive approximations, up to the order indicated at the top of the column, to the first eigenvalues (displayed at the rightmost column) of the operator $A$ for $\alpha=L=1$.}
	\label{asymptotic}
\end{minipage}
\end{table}

The presence of $\log{n}$ terms in $z_n$ can be regarded as the most remarkable effect of the singularity $1/r$. As we will see next, these logarithms are the source of the unusual properties of the spectral functions.

The asymptotic expansion \eqref{asymptotic-eigenvalues} not only shows that the series \eqref{zeta} is convergent for ${\rm Re}(s)>\frac12$, but can also be used to determine the full pole structure of its analytic extension. In order to illustrate the procedure, we replace the first terms of \eqref{asymptotic-eigenvalues} into \eqref{zeta} and expand the resulting expression for large values of $n$,
\begin{align}\label{zeta-zetar}
  \zeta(s)=\frac{(\mu L)^{2s}}{\pi^{2s}}\left\{\zeta_R(2s)
  +\frac{\alpha Ls}{\pi^{2}}
  \left[\zeta'_R(2s+2)
  -(\log{2\pi}+\gamma)\,\zeta_R(2s+2)\right]
  \right\}+\ldots
\end{align}
The ellipsis represents a function which is analytic for ${\rm Re}(s)>-\frac32$; $\zeta_R(s)$ is the Riemann's $\zeta$ function, which has a single pole at $s=1$ with residue $1$.

Representation \eqref{zeta-zetar} shows that $\zeta(s)$ presents a simple pole at $s=\frac12$ with residue $\mu L/2\pi$, proportional to the space volume. Most remarkably, $\zeta(s)$ has a double pole at $s=-\frac12$ with Laurent coefficient $\alpha/8\pi\mu$. The source of this double pole can be traced back to the presence of the logarithm $\log{n}$ in the asymptotic behavior of the eigenvalues shown in \eqref{asymptotic-eigenvalues}.

\bigskip

We summarize the singularities of $\zeta(s)$ up to ${\rm Re}(s)>-\tfrac52$:

\begin{itemize}[leftmargin=*]
	\setlength\itemsep{3mm}
	\item[$\circ$]\quad $s=\frac12$\,: simple pole with residue $\displaystyle{\frac{\mu L}{2\pi}}$ 
	\item[$\circ$]\quad $s=-\frac12$\,: double pole with coefficients $\displaystyle{\frac{\alpha}{8\pi\mu}}$ 
	and $\displaystyle{\frac{\alpha}{4\pi\mu}\left(\log{2\mu L}+\gamma-1\right)}$.
	\item[$\circ$]\quad $s=-\frac32$\,: simple pole with residue $\displaystyle{-\frac{3\alpha^3}{16\pi\mu^3}\left(\frac{\zeta_R(3)}3\,\alpha L+\frac{1}{\alpha L}-\frac{1}{\alpha^2L^2}\right)}$\,.
\end{itemize}

\bigskip

Since the asymptotic expansion of $z_n$ contains increasing powers of $\log{n}$, the function $\zeta(s)$ involves higher derivatives of $\zeta_R(s)$ as one explores regions of the complex plane with lower ${\rm Re}(s)$, so one could expect poles of increasing multiplicity. Nevertheless, an explicit calculation shows that there are cancellations between separate contributions which cause that the remaining singularities of $\zeta(s)$ are just simple poles at negative half-integers. This notwithstanding, there still exist conditions under which poles of increasing multiplicity might appear---this is the subject of Section \ref{higher-poles}.

To clarify this point we follow now a different procedure which explains the absence of poles, other than $s=-\frac12$, with higher multiplicity. Note that this alternative procedure also relies on the asymptotic behavior of expression \eqref{spectrum} for large values of $z_n$.

\begin{figure}[t]
\centering
\begin{minipage}{.7\textwidth}
	\includegraphics[height=20mm]{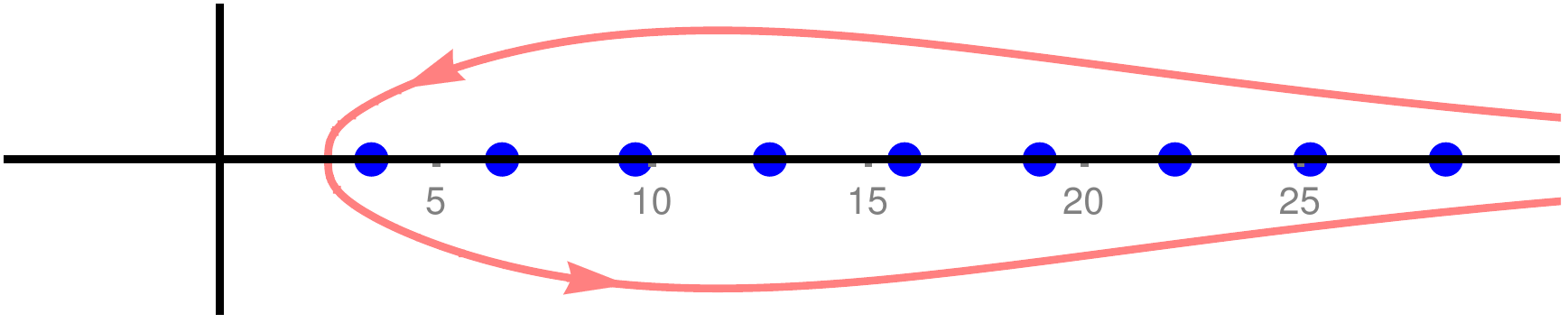}
	\caption{\small{The integration path \curve{$\mathcal C$} enclosing $z_n$ (blue dots), where $z_n^2$ are the positive eigenvalues of $A$. In this figure $\alpha L=1$.}}
	\label{lacurva}
\end{minipage}
\end{figure}

Since the function in the l.h.s.\ of \eqref{spectrum} vanishes at the discrete set $z_n$, the series \eqref{zeta} can also be expressed as
\begin{align}\label{zeta-complex}
  \zeta(s)=\frac{(\mu L)^{2s}}{2\pi i}\oint_{\mathcal{C}}dz\, z^{-2s}
  \ \partial_z\log{\left\{e^{-iz}\ M\left(1+\frac{\alpha L}{2iz},2;2iz\right)\right\}}\,,
\end{align}
where $\mathcal{C}$ is a curve in the complex plane enclosing counterclockwise the positive zeroes $z_n$ of \eqref{spectrum} (see fig.\ \ref{lacurva}). The curve is chosen to lie entirely in the half-plane ${\rm Re}(z)>0$, at finite distance away from $z=0$.
The poles of $\zeta(s)$ arise from the integration at $|z|\rightarrow\infty$.

It is first convenient to turn the complex integration along the imaginary axis. Thus, we write down the asymptotic behaviour of the integrand for $z=\pm i w$ and large $w\in\mathbb{R}^+$,
\begin{align}\label{der-confluent-asymptotic}
  &\partial_w\log{\left\{e^{\pm w}\ M\left(1\mp\frac{\alpha L}{2w},2;\mp2w\right)\right\}}
  \sim  1-\frac1w-\frac{\alpha L}{2w^2}\left(\log{2w}-1\right)+\mbox{}\nonumber\\
  &\mbox{}+\frac{\alpha L}{2w^2}\,\psi\left(1+\frac{\alpha L}{2w}\right)
  +\partial_w\log{\ \mbox{}_2F_0\left(1-\frac{\alpha L}{2w},-\frac{\alpha L}{2w};\frac{1}{2w}\right)}\,.
\end{align}
We have neglected exponentially decreasing terms because, as will become clear below, they do not contribute to the singularities of $\zeta(s)$. The leading term in \eqref{der-confluent-asymptotic} confirms that the integral in \eqref{zeta-complex} is convergent for ${\rm Re}(s)>\frac12$. Expression \eqref{der-confluent-asymptotic} allows us to write \eqref{zeta-complex} as
\begin{align}\label{zeta-complex-asymptotic}
  \zeta(s)&=\frac{(\mu L)^{2s}}{\Gamma(s)\Gamma(1-s)}
  \int_1^\infty dw\, w^{-2s}
  \ \partial_w\log{\left\{e^{-w}\ M\left(1+\frac{\alpha L}{2w},2;2w\right)\right\}}+\ldots
\end{align}
The ellipsis represents an integration along a finite path on the half-plane ${\rm Re}(z)>0$ which runs from $z=i$ to $z=-i$; we disregard this integral for it is a holomorphic function in the whole complex plane $s\in\mathbb C$.

Since the singularities in $\zeta(s)$ arise from the integration at large $w$, we consider the contribution to \eqref{zeta-complex-asymptotic} of each of the terms in \eqref{der-confluent-asymptotic}. The first term in the r.h.s.\ of \eqref{der-confluent-asymptotic} gives
\begin{align}\label{zeta-first}
  \frac{(\mu L)^{2s}}{\Gamma(s)\Gamma(1-s)}
  \int_1^\infty dw\, w^{-2s}=
  \frac{\mu L}{2\pi}\,\frac{1}{s-\frac12}+O(s-\tfrac12)^0\,.
\end{align}
We confirm the simple pole at $s=\frac12$, whose residue---proportional to the volume---coincides with the one due to the leading term in \eqref{asymptotic-eigenvalues}.

The subleading term in \eqref{der-confluent-asymptotic}, proportional to $w^{-1}$, would give a pole at $s=0$ but, because of the denominator $\Gamma(s)$ in \eqref{zeta-complex-asymptotic}, its residue vanishes. Let us remark that $\zeta(s)$ is then regular at $s=0$.

Next, we consider the contribution of the terms in \eqref{der-confluent-asymptotic} proportional to $w^{-2}$ which do not involve $\log{w}$,
\begin{align}\label{s-1}
&-\frac{(\mu L)^{2s}}{\Gamma(s)\Gamma(1-s)}
\int_1^\infty dw\, w^{-2s}
\ \frac{\alpha L}{2w^2}\,(\log{2}-1+\gamma)=\nonumber\\
&=\frac{\alpha}{4\pi\mu}\,(\log{2}+\gamma-1)\,\frac{1}{s+\frac12}+O(s+\tfrac12)^0
\end{align}
More remarkably, the integral of the term involving $\log{w}$ gives
\begin{align}\label{s-1^2}
  &-\frac{(\mu L)^{2s}}{\Gamma(s)\Gamma(1-s)}
  \int_1^\infty dw\, w^{-2s}\ \frac{\alpha L}{2w^2}\,\log{w}=\nonumber\\
  &=\frac{\alpha}{8\pi\mu}\,\frac{1}{\left(s+\frac12\right)^2}
  +\frac{\alpha\log{\mu L}}{4\pi\mu}\,\frac{1}{s+\frac12}
  +O(s+\tfrac12)^0
\end{align}
Collecting \eqref{s-1} and \eqref{s-1^2} we confirm the behavior of $\zeta(s)$ at $s=-\frac12$.

The remaining terms in \eqref{der-confluent-asymptotic} are $O(w^{-3})$ so they give contributions which are analytic for ${\rm Re}(s)>-\tfrac32$. Moreover, they admit an asymptotic expansion in integer powers of $w$ so they do only produce simple poles at negative half-integers (due to the term $\Gamma(s)$ the function $\zeta(s)$ is finite at negative integers).

In conclusion, the logarithm in the asymptotic expansion \eqref{der-confluent-asymptotic} does only introduce a double pole at $s=-\frac12$; the remaining terms, being pure integer powers of $w$, give the usual poles of multiplicity $1$.

\section{The Heat-Trace}\label{theheattrace}

In this section we explore the consequences of the $\log{n}$ behavior of the eigenvalues on the asymptotic expansion of the heat-trace,
\begin{align}\label{htsing}
{\rm Tr}\,e^{-\tau A}
=\sum_{n=1}^\infty e^{-\tau z_n^2}\,.
\end{align}
As an operator on an infinite-dimensional space of functions, the heat-trace diverges as the proper time $\tau$ tends to zero. For smooth backgrounds, this divergent behavior is well-understood through the small-$\tau$ asymptotic expansion \eqref{htexp} in terms of half-integer powers of $\tau$. We now show that the heat-trace of the singular operator \eqref{operator} also contains $\log{\tau}$ terms as $\tau\to0^+$.

If we evaluate \eqref{htsing} using \eqref{asymptotic-eigenvalues} we get
\begin{align}\label{firstheat}
{\rm Tr}\,e^{-\tau A}=\sum_{n=1}^\infty
e^{-\tau\,\frac{\pi^2n^2}{L^2}}
\left\{
1-\frac{\tau\alpha}{L}\left(\log{2\pi n}+\gamma\right)+O(n^{-2},\tau)+O(\tau^2)
\right\}\,.
\end{align}
The first term in the series gives
\begin{align}
\sum_{n=1}^\infty e^{-\tau\,\frac{\pi^2n^2}{L^2}}
\sim \frac{L}{\sqrt{4\pi\tau}}-\frac12\,,
\end{align}
as one expects for the free case $\alpha=0$. The second term can be estimated as
\begin{align}
&-\frac{\tau\alpha}{L}\ \sum_{n=1}^\infty
e^{-\tau\,\frac{\pi^2n^2}{L^2}}\log{n}
-\frac{\tau\alpha}{L}\,(\log{2\pi}+\gamma)\ \sum_{n=1}^\infty
e^{-\tau\,\frac{\pi^2n^2}{L^2}}\nonumber\\[2mm]
&\sim -\frac{\tau\alpha}{L}
\int_0^\infty dn\,
e^{-\tau\,\frac{\pi^2n^2}{L^2}}\log{n}
-\frac{(\log{2\pi}+\gamma)\alpha}{\sqrt{4\pi}}\,\sqrt{\tau}
+O(\tau)\nonumber\\[2mm]
&\sim \frac{\alpha}{4\sqrt{\pi}}\,\sqrt\tau\,\log{\left(\frac{\tau}{L^2}\right)}
-\frac{\gamma\alpha}{4\sqrt{\pi}}\,\sqrt{\tau}
+O(\tau)\,,
\end{align}
where we have used Euler-Maclaurin formula to replace the series by the integral. Collecting these expressions we obtain the small-$\tau$ asymptotics of the heat-trace,
\begin{align}\label{sta}
{\rm Tr}\,e^{-\tau A}\sim \frac{1}{\sqrt{4\pi\tau}}
\left(L
-\sqrt{\pi}\,\tau^{\frac12}
+\frac{\alpha}{2}\,\tau\,\log{\left(\frac{\tau}{L^2}\right)}
-\frac{\gamma\alpha}{2}\,\tau+O(\tau^{\frac32})\right)\,.
\end{align}
Comparing with the asymptotic expansion \eqref{htexp} we get
\begin{align}
	a_0(A)&=L\\
	a_\frac12(A)&=-\sqrt\pi\\
	a_1(A)&=-\alpha\,(\log{L}+\tfrac{\gamma}2)
\end{align}
The coefficient $a_0(A)$ coincides with the usual expression \eqref{a0}. The coefficient $a_\frac12(A)$ also coincides with the case of a smooth background in one dimension under Dirichlet boundary conditions at both endpoints \cite{Vassilevich:2003xt}. On the contrary, $a_1(A)$ is certainly not given by the integral of $1/r$, as in \eqref{a1}. Furthermore, this coefficient contains a $\log{L}$ contribution which suggests---upon dimensional arguments---the presence of another $\log$-term. Indeed, the heat-trace contains a new term $b_1(A)\,\tau\log{\tau}$, not present in the smooth case, with
\begin{align}
	b_1(A)=\frac\alpha2\,.
\end{align}

Using the relation \eqref{mellin} and the asymptotic expansion of the heat-trace \eqref{sta} one confirms that the $\zeta$ function has a simple pole at $s=\frac12$ with residue $\mu L/2\pi$, as well as a double pole at $s=-\frac12$ with Laurent coefficients $\alpha/8\pi\mu$ and $(\log{2\mu L}+\gamma-1)\,\alpha/4\pi\mu$.

\section{The Effective Action}\label{effact}

In this section we will point out some interesting consequences of the double pole in the $\zeta$ function and the $\log$-term in the heat-trace on the effective action computation.

According to \eqref{funder}, one-loop contributions to the effective action of a real massless scalar field in interaction with a Coulomb-type background are given by
\begin{align}
\frac12\,\log{\rm Det}\,D
=-\frac{T}{2}\int_{\Lambda^{-2}}^\infty \frac{d\tau}{\tau}
\ \frac{1}{\sqrt{4\pi \tau}}
\ {\rm Tr}\,e^{-\tau\,A}\,,
\end{align}
where
\begin{align}
	D=-\partial^2_t+A\,,\qquad
	A=-\partial^2_r+\frac{\alpha}{r}\,.
\end{align}
Replacing the asymptotic expansion \eqref{sta} into the heat-trace representation of the determinant we obtain the following UV-divergent terms:
\begin{align}
	-\frac{TL}{8\pi}\,\Lambda^2
	+\frac{T\Lambda}{4\sqrt\pi}
	+\frac {T\alpha}{8\pi}\left(\log{L \Lambda}-\gamma\right)\log{L \Lambda}\,.
\end{align}
Thus, apart from the usual volume and boundary contributions, there is a $\log^2{\Lambda}$ divergence---proportional to $T\alpha$---which should be removed trough an appropriate counterterm.

The $\zeta$-function regularization exhibits a more remarkable aspect. We use \eqref{mellin} to write a relation between the $\zeta$ functions of the operators $D$ and $A$,
\begin{align}\label{zetaad}
	\zeta^{(D)}(s)&=
	\frac{\mu^{2s}}{\Gamma(s)}
	\int_0^\infty d\tau\,\tau^{s-1}
	\ {\rm Tr}\,e^{-\tau(-\partial^2_t+A)}
	=
	\frac{\mu^{2s}}{\Gamma(s)}
	\int_0^\infty d\tau\,\tau^{s-1}
	\,\frac{T}{\sqrt{4\pi\tau}}
	\ {\rm Tr}\,e^{-\tau A}\nonumber\\[2mm]
	&=
	\mu T\ \frac{\Gamma(s-\tfrac12)}{2\sqrt\pi\,\Gamma(s)}
	\ \zeta^{(A)}(s-\tfrac12)
	\,.
\end{align}
As a consequence, the behavior of $\zeta^{(D)}(s)$ at $s=0$ is given by the values of $\zeta^{(A)}(s)$ at $s=-\frac12$. However, the double pole of $\zeta^{(A)}(s-\frac12)$ at $s=0$ is not anymore canceled by $\Gamma(s)$ in the denominator. Therefore, $\zeta^{(D)}(s)$ has a simple pole at $s=0$; this does not jeopardize the computation of the effective action but one must take into account that, in the singular case, the $\zeta$-function approach does not provide a finite renormalization anymore.

The results of Section \ref{seczeta} indicate
\begin{align}\label{zetaa}
	\zeta^{(A)}(-\tfrac12+\epsilon)=
	\frac{\zeta_{-2}}{\epsilon^2}+\frac{\zeta_{-1}}{\epsilon}+\zeta_0+O(\epsilon)\,,
\end{align}
where
\begin{align}
	\zeta_{-2}&=\frac{\alpha}{8\pi\mu}\,,\\[2mm]
	\zeta_{-1}&=\frac{\alpha}{4\pi\mu}
	\left(\log{2\mu L}+\gamma-1\right)\,.
\end{align}
Following \eqref{detzet}, we can now use \eqref{zetaad} and \eqref{zetaa} to compute the functional determinant
\begin{align}
&\log{\rm Det}\left(\frac{D}{\mu^2}\right)
=\left.-\zeta^{(D)}\mbox{}'(\epsilon)\right|_{\epsilon=0}\nonumber\\[2mm]
&=-\frac{T \alpha}{8\pi}
\left(\frac{1}{\epsilon^2}
+4(\log{2}-1)(\log{2\mu L}+\gamma)
-2\log^2{2}-\frac{\pi^2}{6}\right)
+\mu T \zeta_0\,.
\end{align}
Note that the $\zeta$-function regularization produces scale-dependent and divergent terms, proportional to $T \alpha$, which should be absorbed into the classical action. This renormalizes bare parameters associated with the dynamics of the background field---in particular, with terms linear in $V(x)$. Since this procedure strongly depends on the particular dynamics that models the background field we do not pursue this analysis further but focus instead on the remaining contribution $\zeta_0$.

\begin{figure}[t]
	\centering
	\begin{minipage}{.70\textwidth}
		\includegraphics[height=60mm]{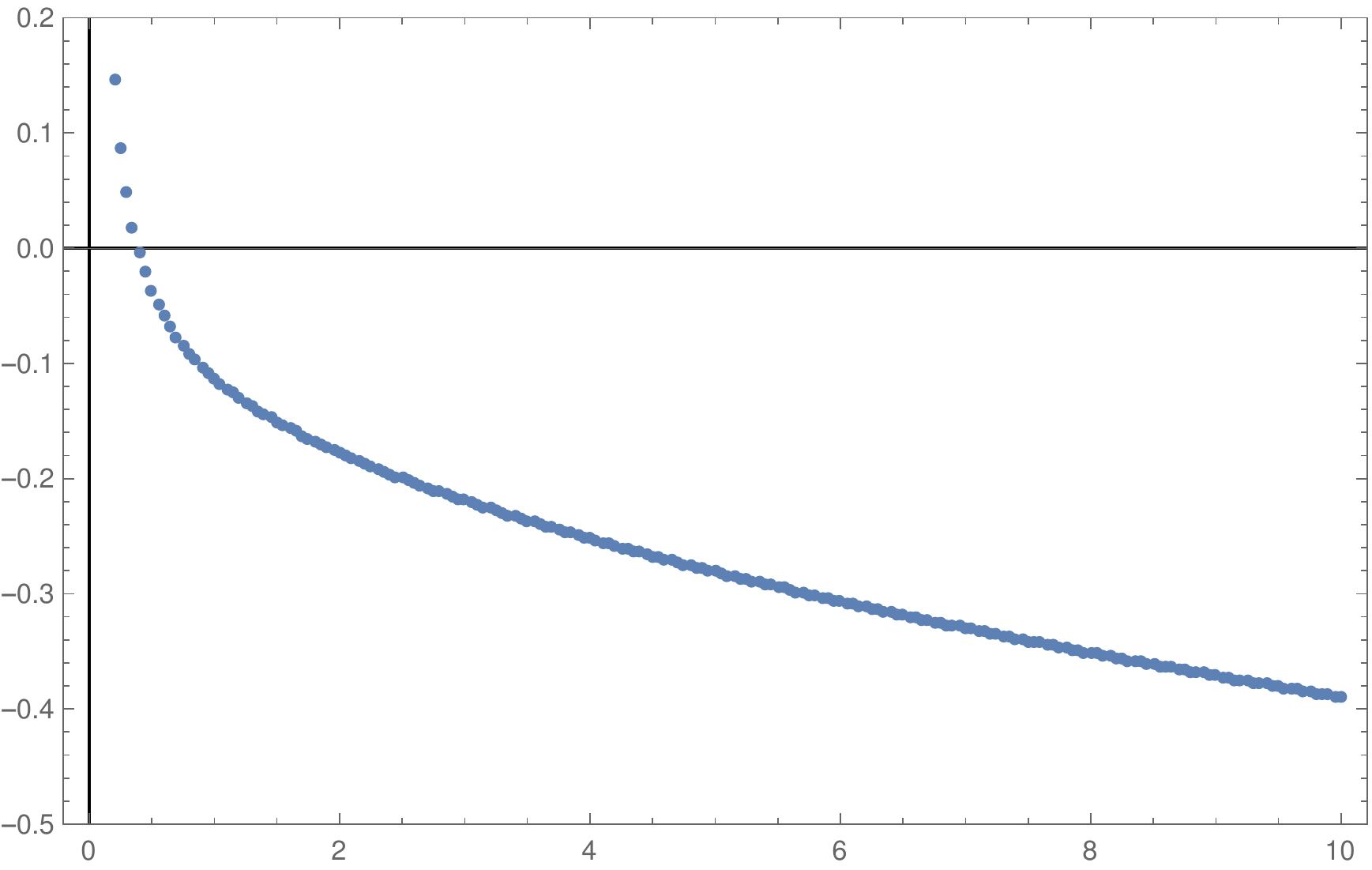}
		\caption{\small{The finite part of the one-loop effective action (per unit time) as a function of $L$, for $\alpha=1$, as represented in the r.h.s.\ of \eqref{logdet}.}}
		\label{zeta0}
	\end{minipage}
\end{figure}

To evaluate $\zeta_0$,  the finite part of $\zeta^{(A)}(s)$ at $s=-\frac12$, we perform the integral in \eqref{zeta-complex} along the imaginary axis. Upon an appropriate rearrangement, and consistently removing terms proportional to $T\alpha$, we obtain
\begin{align}\label{logdet}
	\left.\frac 1T\ \log{\rm Det}\left(\frac{D}{\mu^2}\right)\right|_{\rm reg}
	&=\frac{1}{\pi L}
	\ \Bigg\{
	\frac12-
	\int_0^1 dw\, w
	\ \partial_w\log{\left[e^{-w} M(1+\tfrac{\alpha L}{2w},2;2w)\right]}+\mbox{}\nonumber\\[2mm]
	&\mbox{}-
	\int_1^\infty dw\, w
	\ \bigg(
	\partial_w\log{\left[e^{-w} M(1+\tfrac{\alpha L}{2w},2;2w)\right]}+\mbox{}\nonumber\\[2mm]
	&\mbox{}-\frac{2w^2-2w-\alpha L(\log{2w}+\gamma-1)}{2w^2}\bigg)\Bigg\}
	\,.
\end{align}
Figure \ref{zeta0} shows the r.h.s.\ of this expression as a function of $L$, for $\alpha=1$.

Since the spectrum of $A$ gives the frequencies of the normal modes of a massless field within the rigid boundaries at $r=0$ and $r=L$, the finite part (${\rm FP}$) of $\zeta^{(A)}(s)$ at $s=-\frac12$ represents a contribution---due to the vacuum oscillations of the quantum field---to the ground energy,
\begin{align}
\mathcal E_0=\frac\mu2\ {\rm FP}\,\zeta^{(A)}(-\tfrac12)
=\frac12\ \mu\,\zeta_0\,.
\end{align}
Still, this is a one-loop correction to the contribution given by the background field, which depends on the particular model of the dynamics of $V(x)$.

\section{The Field at the Singularity}\label{bc}

So far we have restricted ourselves to a discussion of the case in which the field vanishes at the singularity, but this is not the most general behavior the singularity admits. In fact, there exist infinitely many boundary conditions under which the singular operator $A$ is selfadjoint. In other words, the standard analysis of its deficiency spaces indicates that $A$ admits a one-parameter family of selfadjoint extensions, each of which corresponds to a particular boundary condition. In appendix \ref{sae} we summarize this approach but in the current section we prefer to follow instead a more intuitive procedure: we analyze the domain $\mathcal D(A)$ of the operator $A$ under some general assumptions on the expected behavior of the fields at $r=0$; this will determine the appropriate boundary conditions.

Let us take $\varphi(r)\in\mathcal D(A)$ and assume $\varphi\sim r^p$ as $r\to0$, for some $p$. Since $\varphi\in \mathbf{L}_2$, then $p>-\frac12$. Moreover, we also expect $\partial_r^2\varphi\in \mathbf{L}_2$, which would further restrict $p>\frac32$. However, this is a too restrictive condition for the fields at $r=0$. Indeed,
\begin{align}
A\,r^p\sim -p(p-1)\,r^{p-2}+\alpha\,r^{p-1}\in \mathbf{L}_2
\end{align}
can also be satisfied for $p\leq \frac32$ as long as $p=1$. Thus, $\varphi(r)\sim r\in\mathcal D(A)$.

It is immediate to see that there exists a linearly independent function which also satisfies $A\,\varphi\in \mathbf{L}_2$, namely $\varphi(r)\sim 1+\alpha r\log{\alpha r}$. This is a consequence of $\partial^2_r(1+\alpha r\log{\alpha r})=\alpha/r$, so that the kinetic terms compensates the non-integrable contribution from the singularity.

We conclude that any function in the domain of $A$ behaves as
\begin{align}\label{phiatzero}
\varphi(r)\sim a_\varphi (1+\alpha r \log{\alpha r})+b_\varphi\,\alpha r+o(r^\frac32)\,,
\end{align}
for some $(a_\varphi,b_\varphi)\in\mathbb{C}^2$. Thus, the ``boundary values'' of each function in $\mathcal D(A)$ determine a point in $\mathbb{C}^2$. On the other hand, selfadjointness implies hermiticity\footnote{For this analysis we omit the contribution of the boundary at $r=L$.},
\begin{align}
0=(A\varphi,\chi)-(\varphi,A\chi)
\sim \varphi'\mbox{}^*\chi-\varphi^*\chi'
\sim b_\varphi^* a_\chi-a_\varphi^*b_\chi\,.
\end{align}
Selfadjoint extensions are given by the Lagrangian subspaces of the boundary-values space $\mathbb{C}^2$ under the symplectic product $b_\varphi^* a_\chi-a_\varphi^*b_\chi$. Therefore, the selfadjoint extensions $A_\beta$ can be characterized by a single parameter $\beta\in\mathbb{R}$, which determines their domain as
\begin{align}\label{domain}
\mathcal D(A_\beta)=\{\varphi(r): b_\varphi=\beta\,a_\varphi\}\,.
\end{align}
Note that $A_\infty$ corresponds to the operator under Dirichlet conditions at $r=0$.

\begin{figure}[t]
	\centering
	\begin{minipage}{.85\textwidth}
		\includegraphics[height=70mm]{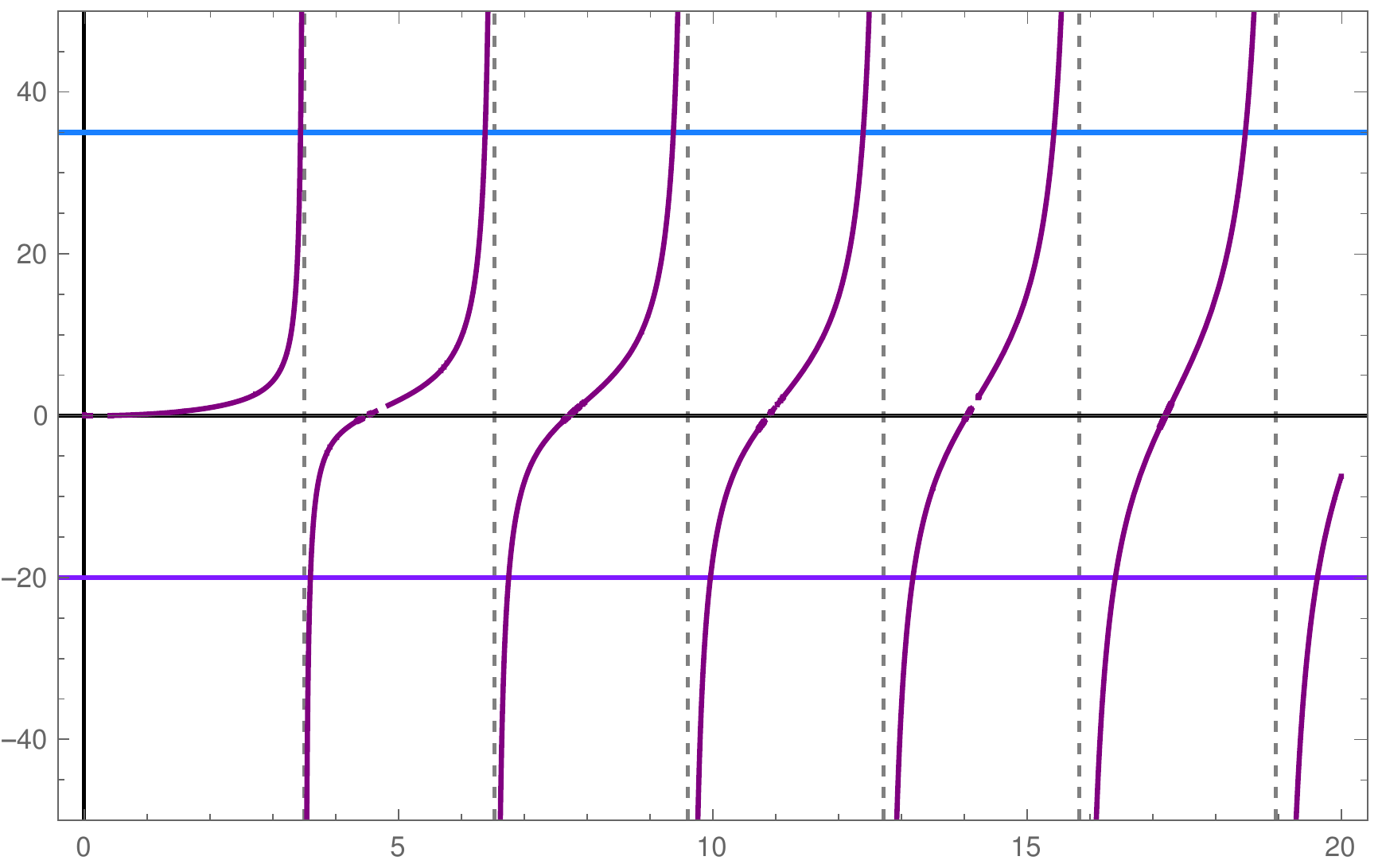}
		\caption{\small{The figure shows the l.h.s.~of \eqref{spec}. The intersections with a given horizontal line at height $\beta$ corresponds to the spectrum of a particular selfadjoint extension $A_\beta$. In the figure we display $\beta=-20$ and $\beta=35$, for $\alpha=1$ (in units of $L$). The dashed asymptotes indicate the spectrum for Dirichlet boundary conditions ($\beta=\infty$).}}
		\label{specbeta}
	\end{minipage}
\end{figure}

Let us now use the boundary condition \eqref{domain} to determine the spectrum of the operator $A_\beta$. Since we are interested in the effects of the singularity $1/r$, we simply impose Dirichlet conditions at $r=L$. Thus, an appropriate solution to \eqref{equation} reads
\begin{align}\label{eigen}
\varphi_z(r)= r\,e^{-iz r}
&\left\{
\ M\left(1+\tfrac{\alpha}{2iz},2;2izL\right)U\left(1+\tfrac{\alpha}{2iz},2;2iz\, r\right)
+\mbox{}\right.\nonumber\\[2mm]
&\left.\mbox{}-
U\left(1+\tfrac{\alpha}{2iz},2;2izL\right)M\left(1+\tfrac{\alpha}{2iz},2;2iz\, r\right)
\right\}\,,
\end{align}
whose behavior close to $r\simeq 0$ is given by
\begin{align}\label{eigenatsing}
\varphi_z(r)&\sim 
M\left(1+\tfrac{\alpha}{2iz},2;2izL\right)
\left(1+\alpha r\log{\alpha r}\right)
+\mbox{}\nonumber\\[2mm]
&\mbox{}+\alpha r
\ \Big\{M\left(1+\tfrac{\alpha}{2iz},2;2izL\right)\left[
\tfrac{iz}{\alpha}+\psi(\tfrac{\alpha}{2iz})+\log{(\tfrac{2iz}{\alpha})}+2\gamma-1
\right]+\mbox{}\nonumber\\[2mm]
&\mbox{}
-U\left(1+\tfrac{\alpha}{2iz},2;2izL\right)\,\Gamma\left(\tfrac{\alpha}{2iz}\right)\Big\}\,.
\end{align}
This behavior is in accordance with our derivation \eqref{phiatzero}. Actually, comparing \eqref{eigenatsing} and \eqref{phiatzero} one readily obtains the equation that determines the spectrum of the operator $A_\beta$:
\begin{align}\label{spec}
f(z)-
\Gamma\left(\tfrac{\alpha}{2iz}\right)\,
\frac{U\left(1+\tfrac{\alpha}{2iz},2;2izL\right)}{M\left(1+\tfrac{\alpha}{2iz},2;2izL\right)}
=\beta\,,
\end{align}
where we have defined the function
\begin{align}\label{efe}
f(z)&:=
\frac{iz}{\alpha}+\psi(\tfrac{\alpha}{2iz})+\log{(\tfrac{2iz}{\alpha})}+2\gamma-1\,.
\end{align}
In figure \ref{specbeta} we plot the l.h.s.\ of \eqref{spec}: The intersections with each horizontal line give the spectrum of a definite selfadjoint extension. The asymptotes ($\beta=\infty$) represent the spectrum for Dirichlet boundary conditions (see e.g.\ fig.\ \ref{lacurva}).

An asymptotic analysis of \eqref{spec} gives the spectrum $z_n^2$ for large $n$,
\begin{align}
	L z_n\sim
	\pi n+\frac\pi2
	-\frac{\alpha L}{2\pi n}
	\left(\log{2\pi n}-2\log{\alpha L}
	+\gamma-2-2\beta\right)+\ldots
\end{align}
Successive approximations would allow us to explore, one by one, the poles of the $\zeta$ function for each selfadjoint extension. However, as in the Dirichlet case, a representation in terms of a contour integration in the complex plane will be much more revealing.

\section{Higher-Order Poles}\label{higher-poles}

In this section we determine the analytic structure of the $\zeta$ function for a particular selfadjoint extension $A_\beta$. In order to do that, we describe its spectrum \eqref{spec} as the zeroes (in units of $L$) of the function
\begin{align}
F_\beta(z)=M\left(1+\tfrac{\alpha L}{2iz},2;2iz\right)
\left(f(\tfrac zL)-\beta\right)
-U\left(1+\tfrac{\alpha L}{2iz},2;2iz\right)\,\Gamma\left(\tfrac{\alpha L}{2iz}\right)\,.
\end{align}
Thus, the $\zeta$ function can be written as
\begin{align}
\zeta_\beta(s)=\frac{(\mu L)^{2s}}{2\pi i}\oint_{\mathcal{C}}dz\, z^{-2s}
\ \partial_z\log{F_\beta(z)}\,,
\end{align}
where $\mathcal C$ encloses the zeroes of $F_\beta(z)$ as in figure \ref{lacurva}. If we deform the integration contour to the imaginary axis we obtain
\begin{align}\label{zeta-imag}
\zeta_\beta(s)=\frac{(\mu L)^{2s}}{2\pi i}
\int_1^\infty dw\, w^{-2s}
\left\{-e^{-i\pi s}\,\partial_w\log{F_\beta(iw)}
+e^{i\pi s}\,\partial_w\log{F_\beta(-iw)}
\right\}+\ldots
\end{align}
where the ellipsis represents a holomorphic function of $s\in\mathbb{C}$. To identify the poles of $\zeta_\beta(s)$ we need the asymptotic expansion of $F_\beta(\pm iw)$ for large $w$,
\begin{align}
	&\partial_w\log{F_\beta(\pm iw)}\sim
	\partial_w\log{M\left(1\mp\tfrac{\alpha L}{2w},2;\mp 2w\right)}+\mbox{}\nonumber\\[2mm]
	&\mbox{}+
	\partial_w\log{\left(
	\frac{w}{\alpha L}-\log{2w}+\log{\alpha L}-\psi(1+\tfrac{\alpha L}{2w})
	-2\gamma+1+\beta
	\right)}\,.
\end{align}
We have explicitly separated the confluent hypergeometric function $M$ because this contribution exactly reproduces the pole structure under Dirichlet conditions, which we have already determined. Consequently, in the rest of this section we will focus on the new poles of $\zeta_\beta(s)$ in relation to those of $\zeta_\infty(s)$.

Replacing the asymptotics of $F_\beta(\pm iw)$ into \eqref{zeta-imag} we obtain
\begin{align}
&\zeta_\beta(s)=\zeta_\infty(s)
+\frac{(\mu L)^{2s}}{\Gamma(s)\Gamma(1-s)}
\int_1^\infty dw\, w^{-2s}\,\times\mbox{}\nonumber\\[2mm]
&\mbox{}\times\,\partial_w\log{\left(1-
	\alpha L\ \frac{\log{2w}-\log{\alpha L}+\psi(1+\tfrac{\alpha L}{2w})
	+2\gamma-1-\beta}{w}\right)}+\ldots\nonumber\\[2mm]
&=\zeta_\infty(s)
+\frac{(\mu L)^{2s}}{\Gamma(s)\Gamma(1-s)}
\int_1^\infty dw\, w^{-2s}
\ \sum_{n=2}^\infty\frac{(\alpha L)^{n-1}}{w^{n}}
\ \sum_{k=0}^{n-1} c_{n,k}\,\log^k{w}+\ldots
\end{align}
where the coefficients $c_{n,k}$ are polynomials in $\beta$, $\alpha L$ and $\log{\alpha L}$. In particular, $c_{n,n-1}=1$. Integration in $w$ gives the pole structure of $\zeta_\beta(s)$,
\begin{align}
\zeta_\beta(s)&=\zeta_\infty(s)+\nonumber\\
&\mbox{}+\frac{(\mu L)^{2s}}{\Gamma(s)\Gamma(1-s)}
\ \sum_{n=2}^\infty\,(\alpha L)^{n-1}
\ \sum_{k=0}^{n-1} c_{n,k}
\ \frac{k!}{2^{k+1}\left(s-\frac{1-n}{2}\right)^{k+1}}+\ldots
\end{align}
In consequence, $\zeta_\beta(s)-\zeta_\infty(s)$ has poles at $s=\frac12-\frac n2$, with $n=2,3,4,\ldots$, of multiplicity $n$ for even $n$, and multiplicity $n-1$ for odd $n$. To illustrate this result we explicitly indicate the behavior of the $\zeta_\beta(s)$ function at its first singularities:
\begin{align}
	s=\tfrac12\ :\quad
	&\frac{\mu L}{2\pi}\,\frac{1}{(s-\tfrac12)}
	+O(s-\tfrac12)^0\nonumber\\[4mm]
	s=-\tfrac12\ :\quad
	&-\frac{\alpha}{4\pi\mu}\,\frac{1}{(s+\tfrac12)^2}
	-\frac{\alpha\,(\log{\tfrac{2\mu}{\tilde\beta\alpha}}-\frac23)}{2\pi\mu}\,\frac{1}{(s+\tfrac12)}
	+O(s+\tfrac12)^0\nonumber\\[4mm]
	s=-1\ :\quad
	&-\frac{\alpha^2}{4\mu^2}\,\frac{1}{(s+1)^2}
	-\frac{\alpha^2\,(\log{\tfrac{2\mu}{\tilde\beta\alpha}}-\frac16)}{2\mu^2}\,\frac{1}{(s+1)}
	+O(s+1)^0\nonumber\\[4mm]
	s=-\tfrac32\ :\quad
	&\frac{3\alpha^3}{8\pi\mu^3}\,\frac{1}{(s+\tfrac32)^4}
	+\frac{3\alpha^3\,\log{\tfrac{2\mu}{\tilde\beta\alpha}}}{4\pi\mu^3}\,\frac{1}{(s+\tfrac32)^3}
	+\mbox{}\nonumber\\[1mm]
	&\mbox{}+
	\frac{3\alpha^3\,(\log^2{\tfrac{2\mu}{\tilde\beta\alpha}}-\frac{1}{9}-\frac{\pi^2}{6})}{4\pi\mu^3}
	\,\frac{1}{(s+\tfrac32)^2}+\mbox{}\nonumber\\[1mm]
	&\mbox{}+
	\frac{\alpha^3\,[\log^3{\tfrac{2\mu}{\tilde\beta\alpha}}
		-(\frac13+\tfrac{\pi^2}2)\log{\tfrac{2\mu}{\tilde\beta\alpha}}
	-\frac{2}{27}-\frac34\zeta_R(3)]}{2\pi\mu^3}\,\frac{1}{\left(s+\tfrac32\right)}+\mbox{}\nonumber\\[2mm]
	&\mbox{}+O(s+\tfrac32)^{0}
\end{align}
To avoid cluttering, we have introduced a new parameter $0<\tilde\beta=e^{\beta-\gamma+\frac43}\leq\infty$.

We conclude that, in the most general case, the singularity $1/r$ generates an infinite series of poles with increasing multiplicities. As opposed to the case of smooth backgrounds (cfr.\ relation \eqref{mellin}), the $\zeta$ function is also singular at negative integers.

\section{Conclusions}\label{conclu}

Inspired by the results that describe the effects of conical singularities on spectral functions, we addressed in this article the corresponding problem for Coulomb-like singularities. We began by considering a one-dimensional Schr\"odinger operator with a Coulomb-type potential under Dirichlet boundary conditions. The most remarkable finding is the presence of a double pole in the meromorphic extension of the $\zeta$ function.

In the context of a quantum field interacting with a static classical background, this $\zeta$ function represents the spacelike part of the operator of quantum fluctuations. Since the addition of each extra dimension makes a shift in the argument of the $\zeta$ function, the timelike part of the operator transforms the double pole at $s=-\frac12$ into a simple pole at $s=0$. In conclusion, the quantity $-\zeta'(0)$ is ill-defined in the 1+1 setting. It is worth stressing that this {\it a priori} unexpected behavior shows up under the simplest boundary condition, namely, the vanishing of the field at the singular point. As a matter of fact, the double pole can be traced back to the presence of a $\log{n}$ term in the subleading behavior of large eigenvalues $\lambda_n$, and this is due to the singular coefficient and not to the type of boundary condition---this is a significant difference with respect to the case of conical singularities. At the same time, one can easily see that this $\log$-term leads to $\log{\tau}$ terms in the small-$\tau$ asymptotics of the heat-trace.

Next, we searched for the most general boundary conditions that make the operator selfadjoint, and found ``hospitable'' conditions---as referred to in \cite{Blau:2006gd}---in the sense that the fields can probe the singularity for they do not need to vanish there. We studied the meromorphic extension of the $\zeta$ function and found, this time, infinitely many multiple poles with increasing multiplicity. At this point it is also important to underline another substantial difference with the case of conical manifolds: in a higher-dimensional setting under $1/r$-type backgrounds, the boundary conditions under which multiple poles appear respect rotational invariance. Indeed, the selfadjoint extensions of the one-dimensional operator $A$ considered in this article apply to each angular momentum eigenspace separately, without mixing them. On the contrary, as shown in \cite{Kirsten:2005yw}, multiple poles for $1/r^2$-type singularities appear for boundary conditions which combine different eigenvalues of the transversal part of the Laplacian (viz., different angular momenta); in particular, they do only appear if the zero modes in the transversal directions play some role in this combination. From this perspective, the multiple poles in the case of $1/r^2$-type backgrounds strongly rely on a peculiarity of the boundary conditions\footnote{It might be interesting to point out that boundary conditions that mix different angular momenta could be related to ``global'' boundary conditions if the tip of the cone were regarded as the limit of circumferences of decreasing radii (strictly speaking, every condition at a single point is, of course, local).}. 

We hope that our results motivate more formal studies on the effects of Coulomb-like singularities on the $\zeta$ function, which would contribute to a better understanding of the behavior of spectral functions when smoothness conditions are relaxed. On the other hand, it would be interesting to pursue a complete analysis of a 3+1-dimensional quantum field in the presence of a Coulomb background. Of course, a specific model for this background should be firstly assumed in order to fully interpret the renormalization of the relevant parameters. Work along this line is currently under consideration.

\section*{Acknowledgments}

The authors would like to thank support from UNLP under grant 11/X748 and from CONICET, Argentina.

\appendix

\section{Selfadjoint Extensions}\label{sae}

In this section we analyze the deficiency subspaces $\mathcal K_\pm={\rm Ker}(A^\dagger\mp i)$ to describe the complete set of selfadjoint extensions of $A$; this determines appropriate boundary conditions. Since we are only interested in the effect of the singularity, we just impose Dirichlet conditions at $r=L$. In this setting, an eigenfunction of $A^\dagger$ can be written as \eqref{eigen} and its behavior at the origin is thus given by \eqref{eigenatsing}. Therefore, the deficiency subspaces are one-dimensional and the selfadjoint extensions are characterized by the isometries $\mathcal U:\mathcal K_+\to \mathcal K_-$, which are thus characterized by a single phase $e^{2i\theta}\in U(1)$.

In this context, the functions in the domain of a particular selfadjoint extension behave near the singularity as
\begin{align}\label{phimas}
\phi(r)\simeq e^{i\theta}\phi_+(r)+e^{-i\theta}\phi_+^*(r)\,,
\end{align}
where $\phi_+(r)$ is an eigenfunction of $A^\dagger$ with eigenvalue $i$ (given by \eqref{eigen} with $z=z_+=\sqrt{i}$).

Finally, a straightforward comparison between an eigenfunction \eqref{eigenatsing} (with $z\in\mathbb{R}^+$) and the general behavior \eqref{phimas} reproduces condition \eqref{domain}, with
\begin{align}
\beta(\theta)
&=\frac{1}{{\rm Re}\left\{e^{i\theta}M(1+\tfrac{\alpha}{2iz_+L},2;2iz_+)\right\}}
\,{\rm Re}\left\{e^{i\theta}
\left(M(1+\tfrac{\alpha}{2iz_+},2;2iz_+L)\right)f(z_+)+\mbox{}\nonumber
\right.\\[2mm]
&\left.\mbox{}
-U(1+\tfrac{\alpha}{2iz_+},2;2iz_+L)\,\Gamma(\tfrac{\alpha}{2iz_+})
\right\}
\,.
\end{align}



\end{document}